  \documentstyle[12pt]{article}
  \def\n{\nonumber}

  \def\th{\theta}

  \def\be{\begin{equation}}
  \def\ee{\end{equation}}
  \def\bq{\begin{eqnarray}}
  \def\eq{\end{eqnarray}}
  
  \def\({\left(}
  \def\){\right)}

\title{A simple shear-free non-singular spherical model with heat flux}
\author{Naresh Dadhich\thanks{E-mail : nkd@iucaa.ernet.in} \\
{\sl Inter-University Centre for Astronomy \& Astrophysics,}\\
{\sl Post Bag 4, Ganeshkhind, Pune - 411 007, India.} \\
L.K. Patel \\
{\sl Department of Mathematics,}\\
{\sl Gujarat University, Ahmedabad 380 009. India}}
 \date{}
 
\begin{document}
 \maketitle

\begin{abstract}
 We obtain an exact simple solution of the Einstein equation describing a 
spherically symmetric cosmological model without the big-bang or any other 
kind of singularity. The matter content of the model is shear free 
isotropic fluid with radial heat flux and it satisfies the weak 
and strong energy conditions. It is pressure gradient combined with 
heat flux that prevents occurrence of singularity. So far all known 
non-singular models have non-zero shear. This is the first shear 
free non-singular model, which is also spherically symmetric.
\end{abstract}
  
  \n PACS numbers: 04.20Jb, 04.60, 98.80Hw
 \newpage 

 Since Senovilla's discovery [1] of an exact singularity free cosmological 
solution of the 
Einstein equation representing a perfect fluid with the equation of state 
$\rho = 3p$ (and subsquently in the same framework the one [2] with 
$\rho = p$), it is now being recognised that
the singularity theorems [3] can not, as generally believed earlier, 
prevent occurrence of non-singular cosmological solutions satisfying all the 
energy and causality conditions. And there is no conflict with the  
theorems in this. The theorems became inapplicable 
because one of the assumptions, existence of closed trapped surface, 
is not respected by these solutions and its violation does not entail any 
unphysical behaviour for the matter content. This assumption was 
however always a suspect [4] but this fact was not fully appreciated 
in absence of a non-singular solution. The Senovilla solution did this 
signal service of dispelling the folklore belief.\\

 A large family of non-singular cosmological models [5] and its 
generalization with heat flux [2] has been considered but they are all 
cylindrically symmetric (see an excellent recent review [6]). For 
practical cosmology, spherical symmetry is however more appropriate. 
It is therefore pertinent to seek spherically symmetric non-singular models. 
The first model of this kind was obtained by one of us [7] which has 
imperfect fluid with heat flux (note, the expression for $\theta$ 
should have a negative sign before it) and it satisfies all the energy 
conditions and has no singularity of any kind. It was obtained by letting one 
of Tolman's solutions [8] expand. The solution has a free time function 
which can be chosen suitably to have non-singular behaviour for physical and 
kinematical parameters and there exist multiple such choices. It is 
also possible to have a non-singular model with null radiation flux 
[9]. These models are both inhomogeneous and anisotropic and have the typical 
behaviour,  beginning with low density at $t\rightarrow -\infty$, 
contracting to high density at $t = 0$ and then again expanding to low 
density as $t\rightarrow \infty$. Nowhere any physical parameter diverges.\\
 
 In the Raychaudhuri equation [10], which governs cosmological dynamics, it is 
acceleration (pressure gradient) and rotation (centrifugal force) that 
counteract gravitational collapse. In cosmology, there is absence of overall 
rotation and hence for checking collapse to avoid singualrity presence 
of acceleration becomes necessary. All the known non-singular 
models [1-2,5,7,9], are not only accelerating but also shearing. 
Though shear acts in favour of collapse in the Raychaudhuri equation 
but its dynamical action through tidal acceleration makes collapse incoherent 
which acts against concentration of large mass in small enough region. 
This would ultimately work against formation of compact trapped 
surface.\\

Raychaudhuri [11] in one of his recent theorems establishes that the 
necessary condition for non-singular cosmological model is that space average 
of physical and kinematical parameters must vanish. That means the 
parameters must depend upon space variables. The space gradient of expansion 
is in vorticity-free spacetime given  by space divergence of shear and heat 
flux [12]. Hence for non-singulaity presence of atleast one of them is 
necessary. The Ruiz-Senovilla family [5] of non-singular cylindrical models is
the example of presence of shear without heat flux. It can be shown that 
presence of shear is in general essential for perfect fluid
G-2 symmetric non-singular 
models [13]. The spherical non-singular model [7] has both shear and heat flux.\\
 
 Then the question arises, could heat flux alone, of course combined 
with pressure gradient, avoid singularity? This is what we wish to demonstrate 
in this letter by obtaining a simple non-singular solution which 
describes an inhomogeneous shear-free spherical model filled with isotropic 
fluid and radial heat flux. The model satisfies the weak and strong energy conditions
as well as has a physically acceptable fall off behaviour in both $r$ and $t$ 
for physical and kinematic parameters. Again there is a free time function which
can be chosen suitably to give non-singular behaviour to model and there 
exist multiple such choices.\\

 The metric of the model is given by

\be
ds^2 = (r^2 + P)^{2n}dt^2 - (r^2 + P)^{2m}[d r^2 + r^2(d \th^2 + 
sin^{2} \th d \phi^2)] \label{1}\\
\ee

\noindent where
$$
2n = 2m \pm \sqrt{8m^2 + 8m +1} \n \\
$$

\noindent in particular
\be
2m = 1 - \sqrt{3/2} < 0, ~2n = \sqrt{3/2}. \label{2} \\
\ee

\noindent Here $P = P(t)$ which can be chosen freely. The Einstein field 
equation for perfect fluid with radial heat flux reads as

\be
R_{ik} - \frac{1}{2}Rg_{ik} = -[(\rho + p)u_iu_k - pg_{ik} + \frac{1}{2} 
(q_iu_k + q_ku_i)] \\ 
\ee

\noindent where we have set $8 \pi G/c^2 = 1$, $u_i u^i = 1 = -q_i q^i, q_i u^i = 0$, 
$\rho, p$ denote fluid density and isotropic pressure, and $q_i$ is 
the radial heat flux vector.

From eqns. (1) and (3) we obtain
\bq
\rho &=& \frac{3m^2{\dot P}^2}{(r^2 + P)^{2n+2}} - 4m 
\frac{3P + (m+1)r^2}{(r^2 + P)^{2m+2}}, \n  \\
p &=& -\frac{m}{(r^2 + P)^{2n+2}}[2(r^2 + P){\ddot P} 
+ (3m-2n-2){\dot P}^2] \n \\
&&+ \frac{4}{(r^2 + P)^{2m+2}}[(m+n)P + n^2r^2], \n \\
q &=& \frac{4m(n+1)r{\dot P}}{(r^2 + P)^{n+2}} 
\eq
\noindent and the expansion and acceleration are given by

\be
\th = \frac{3m{\dot P}}{(r^2 + P)^{n+1}}, ~{\dot u}_r = -\frac{nr}{r^2 + 
P}. \label{7} \\
\ee

We have freedom to choose the function $P(t)$ which could be chosen 
suitably to give non-singular behaviour to the above parameters. As a 
matter of fact there exist multiple choices, for instance $P(t) = a^2 + 
b^2 t^2, a^2 + e^{-b t^2}, a^2 + b^2cos wt, a^2 > b^2$. For all 
these choices it is clear that all the physical and kinematic parameters 
remain regular and finite for the entire range of variables. Note that 
it also admits an interesting oscillating behaviour in time in which the 
model oscillates between two finite regular states. The first case of 
oscillating non-singular model [14] was recently 
considered in the spherical family 
[7]. The oscillating non-singular models are quite novel and interesting of 
their own accord.\\

 In non-oscillating case, all the parameters given above tend to zero 
as $r \rightarrow\infty$ 
or $t\rightarrow\pm\infty$. The universe begins with low density and 
contracts to maximum density and then again expands to low density 
without ever becoming singular. This is a typical behaviour for 
non-singular models [1,2,7]. However in the oscillating case,  
model oscillates in time between two regular finite states, and the 
parameters fall off to zero as $r \rightarrow \infty$. This is how oscillatory  
and non-oscillatory singularity free models differ from each-other in 
their global behaviour.\\ 

 It is obvious from the simple expression for the metric that spacetime 
is causally stable. For verification of the energy contions, we will have 
to find the eigenvalues of the energy momentum tensor, which are given as 
follows:
\be
[\frac{1}{2}(\rho - p + D), ~\frac{1}{2}(\rho - p - D), ~p, ~p], 
D^2 = (\rho + p)^2 - 4q^2.  \\
\ee
   
Note that in all the above expressions there would in view of
eqn. (2) be relative 
dominance of the term of $(r^2 + P)^{-2(m+1)}$. The weak and strong energy 
conditions would require $\rho\ge0, D\ge0, \rho + p + D\ge0, 2p + D\ge0$. 
It can be easily verified that these conditions would hold good for the 
choices for $P(t)$ given above. The dominant energy condition which 
would require $\rho \ge p$ cannot however be satisfied as it is clearly 
violated for large $r$. Thus the model satisfies the weak and strong 
but not the dominant energy condition. \\

 We have thus obatined a spherical model with isotropic pressure fluid 
and radial heat flux without the big-bang or any other kind of singularity. 
This is the first shear free non-singular model. It is inhomogeneous but 
isotropic. It is heat flux that combines with pressure gradient to avoid 
singularity. From the point of view of realistic cosmology, merits of the 
present model are its isotropy and spherical symmetry.\\

 Apart fron the first Senovilla model [1] and a large family of 
cylindrical non-singular models [5,2], there now also exists a large family 
of spherical non-singular models [7,9] and to that the present one 
adds a novel family of shear free models. Even though it does not satisfy 
the dominant energy condition, it is a very simple and interesting model. 
It is remarkable to note that for the first time cosmic singularity has been 
avoided in absence of shear. There has been interesting cases of 
cosmological models [6], for instance [15], which do not satisfy all the 
energy conditions, yet deserve consideration for their other remarkable 
properties. The present model is simple, sheer free and isotropic and
hence is interesting enough. 
Above all it is a very simple spherical model and thus also 
points to an important fact that non-singular 
cosmological solutions are no longer isolated but could occur more 
generally even in spherical symmetry.\\

Acknowledgement: LKP thanks IUCAA for a visit that made this work possible.


\newpage

\begin{thebibliography}{99}
 \bibitem{}
 J.M.M. Senovilla (1990) Phys. Rev. Lett. {\bf 64}, 2219.
 \bibitem{}
 N. Dadhich, L.K. Patel and R. Tikekar (1995) Pramana {\bf 44}, 303.
 \bibitem{}
 S.W. Hawking and G.F.R. Ellis (1973) The large scale structure of 
spacetime(Cambridge University Press).
\bibitem{}
C.W. Misner, K.S. Thorne and J.A. Wheeler (1973) Gravitation (Freeman).
\bibitem{}
E. Ruiz and J.M.M. Senovilla (1992) Phys. Rev. {\bf D45}, 1995.
\bibitem{}
J.M.M. Senovilla (1998) Gen. Relativ. Grav. {\bf 30}, 701. 
\bibitem{}
N. Dadhich (1997) J. Astrophys. Astr. {\bf 18}, 343.
\bibitem{}
R.C. Tolman (1939) Phys. Rev. {bf 55}, 364.
\bibitem{}
N. Dadhich, L.K. Patel and R. Tikekar (1999) Non-singular spherical 
model with null radiation flux, submitted.
\bibitem{}
A.K. Raychaudhuri (1955) Phys. Rev. {\bf 90}, 1123.
\bibitem{}
A.K. Raychaudhuri (1999) Private communication.
\bibitem{}
G.F.R. Ellis (1971) in Proc. International School of Physics "Enrico 
Fermi" XLVII - General Relativity and Cosmology, ed. R. K. Sachs 
(Academic Press, New York). 
\bibitem{}
N. Dadhich and L.K. Patel (1997) Gen. Relativ. Grav. {\bf29}, 179. 
\bibitem{}
N. Dadhich and A.K. Raychaudhuri (1999) Oscillating non-singular 
relativistic spherical model, gr-qc/9901081.
\bibitem{}
M. Mars and J. M. M. Senovilla (1997) Class. Quantum Grav. {\bf17}, 205.

\end{thebibliography}
\end{document}